\def\tipo{2}      
\ifnum \tipo = 2
\documentstyle[pre,times,twocolumn,aps,epsfig]{revtex} 
\def\figsiz1{7cm} \def\frontmatter{\twocolumn[ \hsize\textwidth\columnwidth
\hsize\csname@twocolumnfalse\endcsname}  \else
\documentstyle[preprint,times,aps,prl,epsfig]{revtex} 
\def\figsiz1{12cm} \def\frontmatter{}  \fi

\begin{document}
\draft \frontmatter
\title {STATISTICAL PROPERTIES OF THE INTERBEAT INTERVAL CASCADE IN HUMAN
SUBJECTS}
\author{Fatemeh Ghasemi,$^1$ J. Peinke,$^2$ M. Reza Rahimi Tabar,$^{2,3}$
and Muhammad Sahimi$^4$  }
\address
{\it $^1$Department of Physics, Sharif University of Technology,
Tehran 11365,
Iran\\
$^1$Carl von Ossietzky University, Institute of Physics, D-26111
Oldenburg,
Germany\\
$^3$CNRS UMR 6529, Observatoire de la C$\hat o$te d'Azur, BP 4229,
06304 Nice
Cedex 4, France\\
$^4$Department of Chemical Engineering and Materials Science,
University of Southern California, Los Angeles, California
90089-1211, USA} \maketitle
\bigskip

\maketitle
\begin{abstract}

Statistical properties of interbeat intervals cascade are
evaluated by considering the joint probability distribution
$P(\Delta x_2,\tau_2;\Delta x_1,\tau_1)$ for two interbeat
increments $\Delta x_1$ and $\Delta x_2$ of different time scales
$\tau_1$ and $\tau_2$. We present evidence that the conditional
probability distribution $P(\Delta x_2,\tau_2|\Delta x_1,\tau_1)$
may obey a Chapman-Kolmogorov equation. The corresponding
Kramers-Moyal (KM) coefficients are evaluated. It is shown that
while the first and second KM coefficients, i.e., the drift and
diffusion coefficients, take on well-defined and significant
values, the higher-order coefficients in the KM expansion are very
small. As a result, the joint probability distributions of the
increments in the interbeat intervals obey a Fokker-Planck
equation. The method provides a novel technique for distinguishing
the two classes of subjects in terms of the drift and diffusion
coefficients, which behave differently for two classes of the
subjects, namely, healthy subjects and those with congestive heart
failure.
 \end{abstract}

\pacs{05.10.Gg, 05.40.-a, 05.45.Tp, 87.19.Hh }
\bigskip
\ifnum \tipo = 2 ] \fi

\section{Introduction}
Cardiac interbeat intervals normally fluctuate in a complex
manner.$^{1-6}$ Recent studies reveal that under normal
conditions, beat-to-beat fluctuations in the heart rate may
display extended correlations of the type typically exhibited by
dynamical systems far from equilibrium. It has been argued,$^2$
for example, that the various stages of sleep may be characterized
by long-range correlations of heart rates separated by a large
number of beats. The interbeat fluctuations in the heart rates
belong to a much broader class of many natural, as well as
man-made, phenomena that are characterized by a degree of
stochasticity. Turbulent flows, fluctuations in the stock market
prices, seismic recordings, the internet traffic, and pressure
fluctuations in packed-bed chemical reactors are example of
time-dependent stochastic phenomena, while the surface roughness
of many materials$^{7,8}$ are examples of such phenomena that are
length scale-dependent.

The focus of the present paper is on the intriguing statistical
properties of interbeat interval sequences, the analysis of which
has attracted the attention of researchers from different
disciplines.$^{9-15}$ Analysis of heartbeat fluctuations focused
initially on short-time oscillations associated with breathing,
blood pressure and neuroautonomic control.$^{16,17}$ Studies of
longer heartbeat records, however, revealed $1/f-$like
behavior.$^{18,19}$ Recent analysis of very long time series
indicates that under healthy conditions, interbeat intervals may
exhibit power-law anticorrelations,$^{20}$ follow universal
scaling in their distributions,${21}$ and are characterized by a
broad multifractal spectrum.$^{22}$ Such scaling features change
with the disease and advanced age.$^{23}$ The possible existence
of scale-invariant properties in the seemingly noisy heartbeat
fluctuations is may be attributed to highly complex, nonlinear
mechanisms of physiological control,$^{24}$ as it is known that
circadian rhythms are associated with periodic changes in key
physiological processes.$^{25-33}$ In Figure 1 samples of
interbeats fluctuations of healthy subjects and those with
congestive heart failure (CHF) are shown.
\begin{figure}[c]
\epsfxsize=9truecm\epsfbox{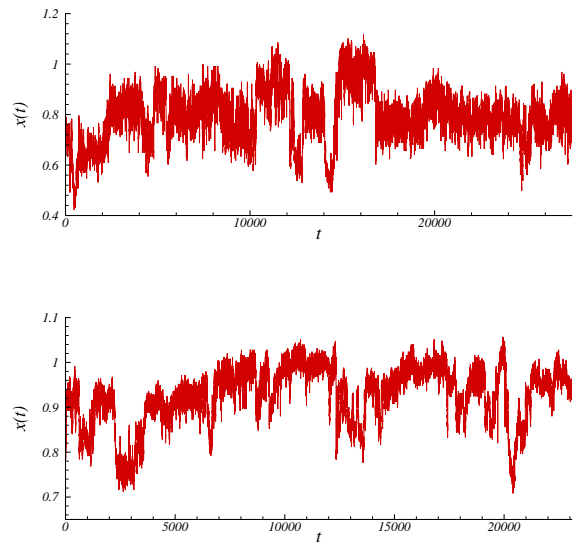} \narrowtext \caption{Time
series of interbeat intervals $x(t)$ versus interval number $t$
for a typical person with congestive heart failure (bottom) and a
healthy subject (top).}
\end{figure}
Recently, Friedrich and Peinke were able$^{34}$ to derive a
Fokker-Planck (FP) equation for describing the evolution of the
probability distribution function of stochastic properties of
turbulent free jets, in terms of the relevant length scale. They
pointed out that the conditional probability density of the {\it
increments}  of a stochastic field (for example, the increments in
the velocity field in turbulent flow) satisfies the
Chapman-Kolmogorov (CK) equation, even though the the velocity
field itself contains long-range, nondecaying correlations. As is
well-known, satisfying the CK equation is a necessary condition
for any fluctuating data to be a Markovian process over the
relevant length (or time) scales.$^{35}$ Hence, one has a way of
analyzing stochastic phenomena in terms of the corresponding FP
and CK equations. In this paper the method proposed by Friedrich
and Peinke is used to compute the Kramers-Moyal (KM) coefficients
for the {\it increments} of interbeat intervals fluctuatations,
$\Delta x(\tau)=x(t+\tau)-x(t)$. Here, $\Delta x$ is the interbeat
increments which, for all the samples, is defined as, $\Delta x
\equiv\Delta x/\sigma_\tau$, where $\sigma_\tau$ is the standard
deviations of the increments in the interbeats data. It is shown
that the first and second KM coefficients representing,
respectively, the drift and diffusion coefficients in the FP
equation, have well-defined values, while the third- and
fourth-order KM coefficients are small. Therefore, a FP evolution
equation$^{35}$ is developed for the probability density function
(PDF) $P(\Delta x,\tau)$ which, in turn, is used to gain
information on changing the shape of PDF as a function of the time
scale $\tau^{36}$ (see also Ref. [37] for another interesting and
carefully-analyzed example of the application of the CK equation
to stochastic phenomena).

The plan of this paper is as follows. In Section 2 we describe the
Friedrich-Peinke method in terms of a KM expansion and the FP
equation. We then apply the method in Section 3 to the analysis of
the increments in the interbeat fluctuations.

\section{The Kramers-Moyal Expansion and Fokker-Planck
Equation}

A complete characterization of the statistical properties of the
interbeat fluctuation requires evaluation of the joint PDFs,
$P_N(\Delta x_1, \tau_1,\cdots,\Delta x_N,\tau_N)$, for an
arbitrary $N$, the number of data points. If the phenomenon is a
Markov process, an important simplification arises in that, the
$N$-point joint PDF $P_N$ is generated by the product of the
conditional probabilities $P(\Delta x_{i+1},\tau_{i+1}|\Delta
x_i,\tau_i)$, for $i=1,\cdots,N-1$. Thus, as the first step of
analyzing a stochastic time series, we check whether the
increments in the data follow a Markov chain. As mentioned above,
a necessary condition for a stochastic phenomenon to be a Markov
process is that the CK equation,$^{34}$
\begin{eqnarray}
& &P(\Delta x_2, \tau_2|\Delta x_1,\tau_1)= \cr \nonumber\\
& &\int d(\Delta x_3)\; P(\Delta x_2, \tau_2|\Delta
x_3,\tau_3)\;P(\Delta x_3,\tau_3|\Delta x_1, \tau_1)\;,
\end{eqnarray}

should hold for any value of $\tau_3$, in the interval
$\tau_2<\tau_3< \tau_1.^{35}$ Therefore, we check the validity of
the CK equation for describing the data using many values of the
$\Delta x_1$ triplets, by comparing the directly-evaluated
conditional probability distributions $P(\Delta x_2,\tau_2|\Delta
x_1,\tau_1)$ with those calculated according to right-hand side of
Eq. (1). In Fig. 2, the directly-computed PDF is compared with the
one obtained from Eq. (1). Allowing for a statistical error of the
order of the square root of the number of events in each bin, we
find that the PDFs are statistically identical.

\begin{figure}[c]
\epsfxsize=9truecm\epsfbox{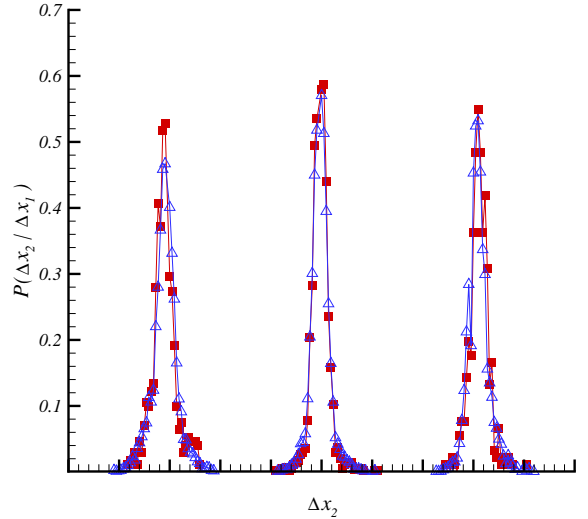}
 \narrowtext \caption{Test of Chapman-Kolmogorov equation for $\Delta x_1=-0.42
$, $\Delta x_1=0$ and $\Delta x_1=0.42$. The solid and open
symbols represent, respectively, the directly-evaluated PDF and
the one obtained from Eq. (1). The PDFs are shifted in the
horizontal directions for clarity. Values of $\Delta x$ are
measured in units of the standard deviation of the increments. The
time scales $\tau_1$, $\tau_2$ and $\tau_3$ are $10$, $30$, and
$20$, respectively.}
\end{figure}
It is well-known that the CK equation yields an evolution equation
for the distribution function $P(\Delta x,\tau)$ across the scales
$\tau$. The CK equation, when formulated in differential form,
yields a master equation, which takes on the form of a FP
equation:$^{35}$
\begin{eqnarray}
& &\frac {d}{d\tau}P(\Delta x,\tau)=\cr \nonumber\\
& &\left[-\frac{\partial }{\partial\Delta x} D^{(1)}(\Delta
x,\tau) +\frac{\partial^2}{\partial\Delta x^2}D^{(2)}(\Delta
x,\tau)\right]P(\Delta x,\tau)\;.
\end{eqnarray}
The drift and diffusion coefficients, $D^{(1)}(\Delta x,\tau)$ and
$D^{(2)}(\Delta x,\tau)$, are estimated directly from the data and
the moments $M^{(k)}$ of the conditional probability
distributions:
\begin{equation}
D^{(k)}(\Delta x,\tau)=\frac{1}{k!}\lim_{\Delta\tau\to
0}M^{(k)}\;,
\end{equation}
\begin{equation}
M^{(k)}=\frac{1}{\Delta \tau}\int d\Delta x'\;(\Delta
x'-\Delta\tau)^k P(\Delta x',\tau+\Delta\tau|\Delta x,\tau)\;.
\end{equation}
The coefficients $D^{(k)}(\Delta x,\tau)$ are known as the
Kramers-Moyal (KM) coefficients.

\subsection{Application to Analyzing Heartbeat Data}

As an application of the method, we analyzed both daytime (12:00
pm to 18:00 pm) and nighttime (12:00 am to 6:00 am) heartbeat time
series of healthy subjects, and the daytime records of patients
with CHF. Our data base includes 10 healthy subjects (7 females
and 3 males with ages between 20 and 50, and an average age of
34.3 years), and 12 subjects with CHF, with 3 females and 9 males
with ages between 22 and 71, and an average age of 60.8 years).
The resulting drift and diffusion coefficients, $D^{(1)}$ and
$D^{(2)}$, are displayed in Figures 3 and 4. It turns out that the
drift coefficient $D^{(1)}$ is a linear function of $\Delta x$,
whereas the diffusivity $D^{(2)}$ is quadratic in $\Delta x$.
Estimates of these coefficients are less accurate for large values
of $\Delta x$ and, thus, the uncertainties increase. Using the
data set for the healthy subjects we find that,
\begin{figure}[c]
\epsfxsize=9truecm\epsfbox{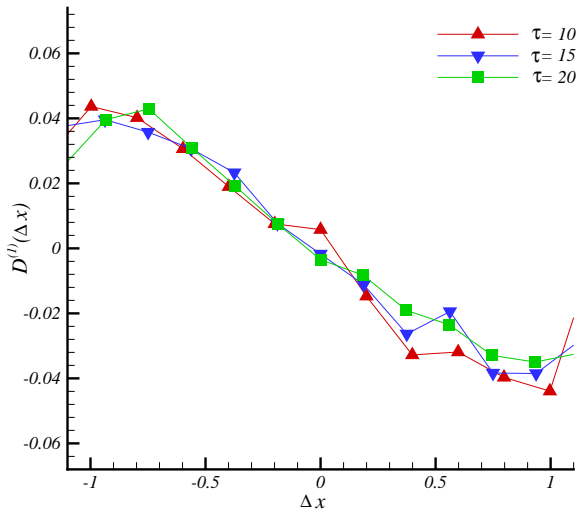}
\epsfxsize=9truecm\epsfbox{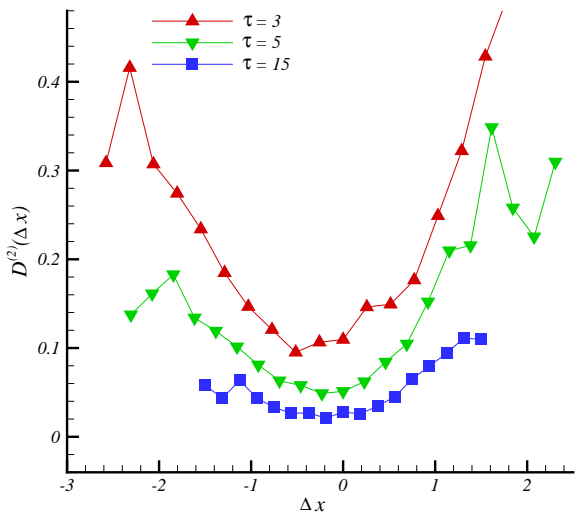}
 \narrowtext\caption{The drift
and diffusion coefficients $D^{(1)}(\Delta x)$ and $D^{(2)}(\Delta
x)$, estimated from Eq. (5) for a healthy subject, follow linear
and quadratic behavior, respectively.}
\end{figure}
\begin{eqnarray}
& & D^{(1)} (\Delta x,\tau)=-0.03\Delta x-0.0046\;,\cr \nonumber\\
& & D^{(2)}(\Delta x,\tau)=\cr \nonumber\\
& &\left(0.01+\frac{0.11}{\tau}\right)(\Delta
x)^2+\left(0.057+\frac{0.287}{\tau}\right)\;,
\end{eqnarray}
whereas for the patients with CHF we obtain,
\begin{eqnarray}
& & D^{(1)}(\Delta x,\tau)=-0.013\Delta x-0.0018\;,\cr\nonumber\\
& & D^{(2)}(\Delta x,\tau)=\cr \nonumber\\
& &\left(0.005+\frac{0.005}{\tau}\right)(\Delta x)^2+
\left(0.013+\frac{0.066}{\tau}\right)\;.
\end{eqnarray}

We also computed the {\it average} of the coefficients $D^{(1)}$
and $D^{(2)}$ for the entire set of the healthy subjects, as well
as those with CHF. According to the Pawula`s theorem,$^{34,37}$
the KM expansion is truncated after the second term, provided that
the fourth-order coefficient $D^{(4)}(\Delta x,\tau)$ vanishes.
For the data that we analyze the coefficient $D^{(4)}$ is about
$\frac{1}{10}D^{(2)}$ for the healthy subjects, and about
$\frac{1}{20}D^{(2)}$ for those with CHF.

Equations (5) and (6) state that the drift coefficients for the
healthy subjects and those with CHF have the same order of
magnitude, whereas the diffusion coefficients for given $\tau$ and
$\Delta x$ are different by about one order of magnitude. This
points to a relatively simple way of distinguishing the two
classes of the subjects. Moreover, the $\tau$-dependence of the
diffusion coefficient for the healthy subjects is stronger than
that of those with CHF (in the sense that the numerical
coefficients of the $\tau^{-1}$ are larger for the healthy
subjects). These are shown in Figures 3 and 4.

The strong $\tau-$dependence of the diffusion coefficient
$D^{(2)}$ for the healthy subjects indicates that the nature of
the PDF of their increments $\Delta x$ for given $\tau$, i.e.,
$P(\Delta x,\tau)$, is intermittent, and that its shape should
change strongly with $\tau$. However, for the subjects with CHF
the PDF is not so sensitive to the change of the time scale
$\tau$, hence indicating that the increments' fluctuations for the
subjects with CHF is {\it not} intermittent. These results are
completely compatible with the recent discoveries that the
interbeat fluctuations for healthy subjects and those with CHF
have fractal and multifractal properties, respectively.$^{22}$

\section{Summary} \label{detcross}
We have shown that the probability density of the interbeat
interval increments satisfies a Fokker-Planck equation, which
encodes the Markovian nature of the increments' fluctuations. We
have been able to compute reliably the first two Kramers-Moyal
coefficients for the stochastic processes $\Delta x$ - the drift
and diffusion coefficients in the FP representation - and, using
the polynomial ansatz,$^{34}$ obtain simple expressions for them
in terms of $\Delta x$ and the time scale $\tau$. We have shown
that the drift and diffusion coefficients of the increments in the
interbeat fluctuations of healthy subjects and patients with CHF
have different behavior, when analyzed by the method we use in
this paper. Hence, they help one to distinguish the two groups of
the subjects. Moreover, one can obtain the form of the path
probability functional of the increments in the interbeat
intervals in the time scale, which naturally encodes the scale
dependence of the probability density. This, in turn, provides a
clear physical picture of the intermittent nature of interbeat
intervals fluctuations.
\begin{figure}[c]
\epsfxsize=9truecm \epsfbox{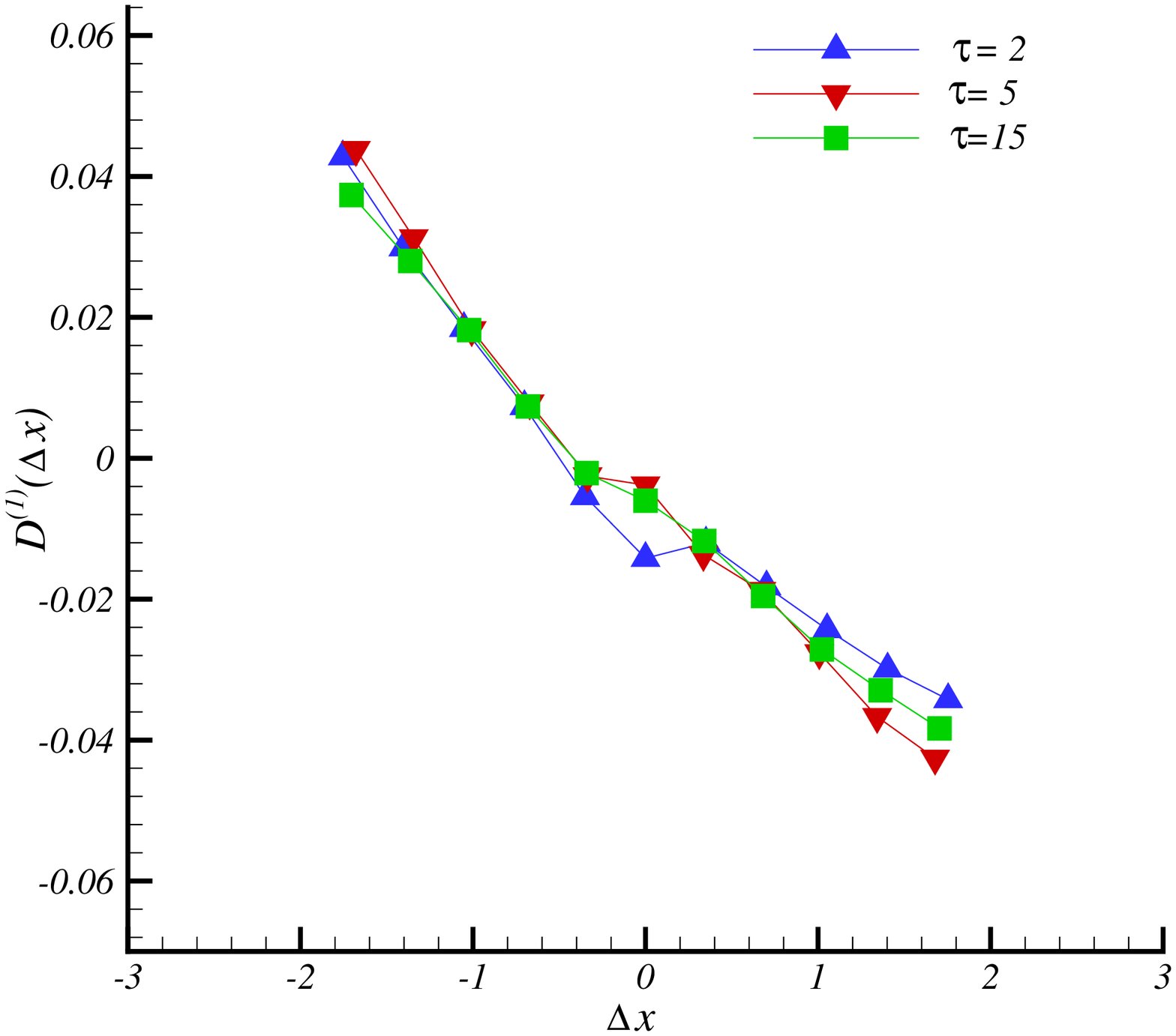} \epsfxsize=9truecm
\epsfbox{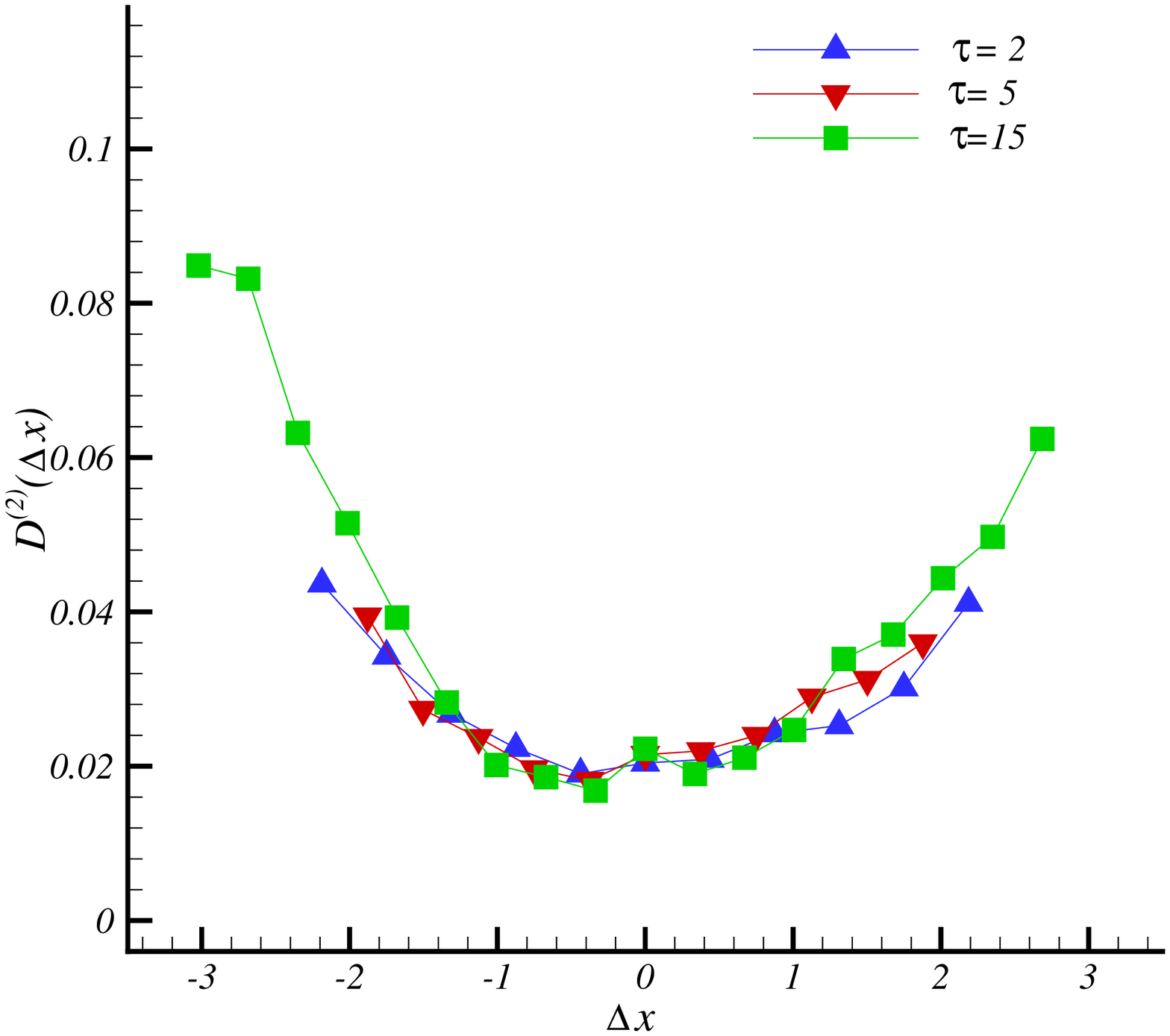}
 \narrowtext \caption{The
drift and diffusion coefficients $D^{(1)}( \Delta x )$ and
$D^{(2)}(\Delta x)$ are estimated from the Eq. (6) for typical
patients with congestive heart failure, and follow linear and
quadratic behavior, respectively.}
\end{figure}
Let us emphasize that the previous analysis$^{1-6}$ of the data
that we consider in this paper indicated that there may be
long-range correlations in the data which might be characterized
by self-affine fractal distributions, such as the fractional
Brownian motion or other types of stochastic processes that give
rise to such correlations. In that method one distinguishes
healthy subjects from those with CHF in terms of the {\it type} of
the correlations that might exist in the data. For example, if the
data follow a fractional Brownian motion, then the corresponding
Hurst exponent $H$ is used to distinguish the two classes of the
subjects, as $H<0.5$ ($>0.5$) indicates negative (positive)
correlations in the data, while $H=0.5$ indicates that the
increments in the data follow Brownian motion. The method proposed
in the present paper is different from such analyses in that, the
{\it increments} in the data are analyzed in terms of Markov
processes. This is {\it not} in contradiction with the previous
analyses. Our analysis does indicate the existence of correlations
in the increments, but, as is well-known in the theory of Markov
processes, such correlations, though extended, eventually decay.
We distinguish the healthy subjects from those with CHF in terms
of the {\it differences} between the drift and diffusion
coefficients of the Fokker-Plank equation that we construct for
the incremental data which, in our view, provides a clearer and
more physical way of understanding the differences between the two
groups of the subjects than the previous method.

\end{document}